\documentstyle[12pt]{article}

\author{S. A. Dias and M. B. Silva Neto\thanks{
e-mail: tiao@cbpfsu1.cat.cbpf.br 
and silvanet@cbpfsu1.cat.cbpf.br}\\
Centro Brasileiro de Pesquisas F\'\i sicas - CBPF\\
Departamento de Teoria de Campos e Part\'\i culas - DCP\\
Rua Dr. Xavier Sigaud, 150 - Urca\\
22290-180 - Rio de Janeiro, RJ - Brazil}
\title{On Bosonization Ambiguities of Two Dimensional Quantum  
Electrodynamics
}
\date{{ }
}

\begin{document}

\maketitle
\begin{abstract}
We study bosonization ambiguities in two dimensional quantum 
electrodynamics
in the presence and in the absence of topologically charged 
gauge fields.
The computation of fermionic correlation functions suggests that 
ambiguities
may be absent in nontrivial topologies, provided that we do not allow
changes of sector as we evaluate functional integrals. 
This would remove an
infinite arbitrariness from the theory. In the case of trivial
 topologies,
we find upper and lower bounds for the Jackiw-Rajaraman parameter,
corresponding to the limiting cases of regularizations which
 preserve gauge
or chiral symmetry.
\end{abstract}

\newpage 

\section{Introduction}

The study of models in dimensions other than four frequently
 brings out
features that give insights about what happens in higher dimensions.
 In two
dimensions, the procedure known as bosonization \cite{colleman}\cite
{mandelstam} sometimes gives a gaussian expression for the functional
integral, making it possible to obtain its exact solution and then to
compare the results with those given by perturbative techniques. 
This allows
the explicit study of characteristics such as charge screening and
 vacuum
structure \cite{schwinger}\cite{thirring}\cite{lowenstein-swieca}
 which are
very important for theories like QCD$_4$.

A characteristic feature of these models is the appearance of
 anomalies
after the regularization of some ill defined quantities.
 This can be easily
seen from the cohomological point of view
 \cite{zumino}\cite{stora}\cite
{bonora}. Let $\Gamma $ be the effective action of, 
for example, QED$_2$
with Dirac fermions (the so called {\it Schwinger model}).
 This theory has
classical gauge and chiral symmetries. We can construct,
 using appropriate
ghost fields, two nilpotent BRST operators
 $\Sigma _g$ and $\Sigma _{ch}$
associated respectively to each one of them.
 The action of these operators
over $\Gamma $ gives, in general%
$$
\Sigma _g\Gamma =\Delta _g, 
$$
$$
\Sigma _{ch}\Gamma =\Delta _{ch}, 
$$
and the following consistency conditions have to be satisfied%
$$
\Sigma _g\Delta _g=0, 
$$
\begin{equation}
\label{2problemas}\Sigma _{ch}\Delta _{ch}=0. 
\end{equation}
The effective action would preserve both 
symmetries at quantum level if the
condition 
\begin{equation}
\label{cohomologia}\left( \Sigma _g+\Sigma _{ch}\right) \Gamma =0 
\end{equation}
were satisfied. Due to the fact that 
$$
\left( \Sigma _g+\Sigma _{ch}\right) ^2=0, 
$$
we can see that condition (\ref{cohomologia}) will be satisfied for
$\Gamma ^{\prime }=\Gamma - C$ and 
$$
\Delta _g+\Delta _{ch}=\left( \Sigma _g+\Sigma _{ch}\right) C, 
$$
where $C$ is a local polynomial in the fields and their derivatives,
 not
involving ghost fields. There is no such $C$ for the case under
 consideration.
This implies the existence of an anomaly, which is a
non-trivial cocycle of the composite operator
 $\Sigma _g+\Sigma _{ch}$.

However, if we solve the two distinct cohomology problems
 (\ref{2problemas})
we find 
$$
\Delta _g=\Sigma _gC_g, 
$$
$$
\Delta _{ch}=\Sigma _{ch}C_{ch}. 
$$
This garantees that we can choose which symmetry to preserve.
 For example,
if we want to preserve chiral symmetry, we redefine 
$\Gamma $ as (it is an
addiction of local counterterms)%
$$
\Gamma ^{\prime }=\Gamma -C_{ch} 
$$
and obtain $\Sigma _{ch}\Gamma ^{\prime }=0$ but 
$\Sigma _g\Gamma ^{\prime
}\neq 0$.

The presence of this anomaly can be seen if one regularizes
 QED$_2$ in a
generalized way \cite{jackiw-johnson}, 
without necessarily preserving gauge
symmetry, in which case one finds for 
the divergencies of the gauge and
axial currents%
$$
\left\langle \partial _\mu J^\mu 
\right\rangle =\frac e\pi \partial _\mu
A^\mu \left( 1-a\right) , 
$$
$$
\left\langle \partial _\mu J_5^\mu 
\right\rangle =-\frac e{2\pi }\epsilon
^{\mu \nu }F_{\mu \nu }a, 
$$
where bosonization introduced the arbitrary parameter $a$.
 The parameter $a$
manifests itself in the position of the pole
 in the photon propagator. This
position is a physical observable \cite{gross}, 
the mass of the gauge boson,
dynamically generated. On the other hand, 
in the input lagrangean, we have
only one parameter to be fixed by 
''experimental data'', the charge $e$ of
the fermion. The new parameter is thus 
completely arbitrary, giving us the
impression that we ended up with ill 
defined predictions after the end of
the quantization procedure.

Thus, we see ourselves facing this question: 
are there physical parameters
of a quantum field theory to which we 
have no classical access? In the case
of the Schwinger model this question is 
not usually asked because the value $a=1$
preserves gauge invariance at quantum level, 
providing the easiest approach
to this model. However, all values of $a$ 
seem, at first sight, to define a
healthy theory. In fact, there is an 
equivalence between the Schwinger
model, for any value of $a$ and a theory 
obtained by the addition of an
extra quantum degree of freedom, with 
action given by the Wess-Zumino action 
\cite{faddeev}. In this context, Harada 
and Tsutsui \cite{harada} obtained a
quantum gauge invariant theory for 
{\it any value of the ambiguous parameter}%
, thus showing that an anomalous theory 
can be viewed as a kind of gauge
fixed version of a gauge invariant theory 
with more degrees of freedom.
Harada and Tsutsui's formalism shows that 
a gauge anomaly is not an obstacle
to consistently quantize a field theory, 
at least in two dimensions. This
point of view sugests that there is no 
naturally favoured value for $a$. We
should study this theory for other 
values of $a$ in order to decide if
different values give different physical implications or not.

In a previous work \cite{tiao-cesar}, 
we have noticed that the computation
of correlation functions in nontrivial topology sectors could give a
mathematical criterium to fix the value of 
$a$ for each given sector (except
the trivial one). This has raisen the hope 
that, perhaps with a mix between
physical requirements and mathematical 
skill, one could argue in favour of a
fixed value for $a$. So, we decided 
to face this question in the context of
the Schwinger model, which is very well 
known for $a=1$, and to see if this
value is favoured by arguments other than gauge invariance.

The paper is organized as follows: in 
section 2 we briefly review QED$_2$ in
the general case where the gauge field 
can be given a topological charge; in
section 3 we compute the contributions 
of these nontrivial sectors to
correlation functions with general $a$ 
and give an argument to fix its value
in all these sectors; in section 4 we 
perform the same analysis in the
trivial topology sector and find restrictions, based on physical
requirements, in the range of values that 
$a$ can assume. Finally, in
section 5, we present our conclusions and some remarks.

\section{The model}

We will study quantum electrodynamics 
in two dimensional euclidean space
described by the action functional 
\begin{equation}
\label{lagrangean}S=\int d^2x{\cal L}
\left( A_{\mu},\overline{\psi }, \psi
\right) =\int d^2x\left[ 
\frac 14F_{\mu \nu }F^{\mu \nu }+\overline{ \psi }%
D\psi \right] , 
\end{equation}
where $D$ is the Dirac operator 
\begin{equation}
\label{dirac-operator}D\equiv 
\gamma ^\mu \left( i\partial _\mu +eA_\mu
\right) . 
\end{equation}
Our $\gamma $ matrices satisfy 
$$
\left\{ \gamma _\mu ,\gamma _\nu \right\} 
=2\delta _{\mu \nu },\qquad \gamma
_5=i\gamma _0\gamma _1,\qquad \gamma _\mu ^{\dagger }=\gamma _\mu ,
$$
which implies, in two dimensions, 
$$
\gamma _\mu \gamma _5=i\epsilon _{\mu \nu }\gamma _\nu , 
$$
where $\epsilon $$_{01}=-\epsilon _{10}=1.$

The generating functional of correlation 
functions for the Schwinger model
is given by 
\begin{equation}
\label{generating-functional}Z\left[
 J^\mu ,\overline{\eta },\eta \right]
=\int \left[ dA_\mu \right] \left[ 
d\overline{\psi }\right] \left[ d\psi
\right] \exp \left( -S+\left\langle J^\mu 
A_\mu \right\rangle +\left\langle 
\overline{\eta }\psi \right\rangle +
\left\langle \overline{\psi }\eta
\right\rangle \right) , 
\end{equation}
where $J^\mu ,\overline{\eta }$ and $\eta $ are the external sources
associated with the fields $A_\mu ,\psi $ and $\overline{\psi },$
respectively. In order to define the functional measure in (\ref
{generating-functional}), we write the fermionic fields as linear
combinations 
\begin{eqnarray}
\psi \left( x\right) &=&\sum_na_n\varphi _n\left( x\right)
+\sum_ia_{0i}\varphi _{0i}\left( x\right) , \\
\overline{\psi }\left( x\right) &=&\sum_n\overline{a}_n\varphi _n^{
\dagger
}\left( x\right) +\sum_i\overline{a}_{0i}\varphi _{0i}^{\dagger }
\left(
x\right) ,
\end{eqnarray}
of the eigenfunctions of $D$ 
\begin{eqnarray}
D\left( A_\mu \right) \varphi _n\left( x\right) &=&\lambda _n\varphi
_n\left( x\right) ,  \label{eingenvalue eq} \\
D\left( A_\mu \right) \varphi _{0i}\left( x\right) &=&0,
\end{eqnarray}
with $a_n$, $\overline{a}_n$ and $a_{0i}$, $\overline{a}_{0i}$ being
grassmanian coeficients. Now, 
the fermionic functional measure is simply 
$$
\left[ d\overline{\psi }\right] 
\left[ d\psi \right] =\prod_nd\overline{a}%
_nda_n\prod_id\overline{a}_{0i}da_{0i}\textstyle{,} 
$$
such that, after an integration over 
fermi fields, the fermionic part of the
generating functional can be written as 
$$
Z_F\left[ \overline{\eta },\eta \right] 
\propto \det {}^{\prime }D\textstyle{%
.} 
$$
In the above expression, det$^{\prime }D$ 
stands for the product of all
nonvanishing eigenvalues of $D$.

As it is well known \cite{abdalla-abdalla-rothe}, 
the appearance of $N$ zero
eigenvalues associated to the Dirac 
operator, the zero modes, is closely
related to the existence of classical 
configurations, in the gauge field
sector, which can be written as \cite{bardakci-crescimano}\cite
{manias-naon-trobo} 
$$
eA_\mu ^{\left( N\right) }=-\tilde \partial _\mu f, 
$$
where $\tilde \partial _\mu \equiv 
\epsilon _{\mu \nu }\partial _\nu$ and
the function $f\left( x\right) $ behaves, at infinity, as 
$$
\lim _{\left| x\right| \rightarrow 
\infty }f\left( x\right) \simeq -N\ln
\left| x\right| . 
$$
These configurations carry a topological 
charge $Q=N$, where $Q$ is given by 
$$
Q=\frac 1{4\pi }\int d^2x\epsilon _{\mu 
\nu }F_{\mu \nu } \textstyle{.} 
$$
For future purposes, we will define 
$$
A_\mu ^\alpha =A_\mu ^{\left( N\right) }+\alpha a_\mu \textstyle{,} 
$$
where $\alpha $ is an interpolating 
parameter \cite{tiao-cesar} between the
fixed configuration 
$A_\mu ^{\left( N\right) }$ ($\alpha =0$) and a general
configuration with topological charge 
$N$ ($\alpha =1$). In two dimensions
we can always write any configuration 
of charge $N$ linearly in terms of a
fixed one, due to the additive property 
of the topological charge. The field 
$a_\mu $ has vanishing topological 
charge and can always be written as 
\begin{equation}
\label{ro-fi}ea_\mu =\partial _\mu 
\rho -\tilde \partial _\mu \phi \textstyle%
{.} 
\end{equation}

The Dirac operator 
$$
D_\alpha =\gamma ^\mu \left( i\partial _\mu +e\left( A_\mu ^{ \left(
N\right) }+\alpha a_\mu \right) \right) , 
$$
has an inverse only if we add a small mass $\epsilon >$$0$, 
$$
\left( D_\alpha +\epsilon 
{\bf 1}\right) ^{-1}\left( x,y\right) ={\ {\bf S}}%
_\epsilon ^\alpha \left( x,y\right) +\frac 1\epsilon 
{\ {\bf P}}_0^\alpha
\left( x,y\right) , 
$$
where $P_0^\alpha \left( x,y\right) $ 
is the projector on the subspace
generated by the zero modes of $D_\alpha $ 
\begin{equation}
\label{proj-md-zero}{\bf P}_0^\alpha 
\left( x,y\right) =\sum_{i=1}^{ \left|
N\right| }\varphi _{0i}^\alpha \left( x\right) \varphi _{0i}^{\alpha
^{\dagger }}\left( y\right) , 
\end{equation}
and ${\bf S}_\epsilon ^\alpha \left( x,y\right) $ inverts $D_\alpha
+\epsilon {\bf 1}$ in the rest of the space and is formally given by 
$$
{\bf S}_\epsilon ^\alpha 
\left( x,y\right) =\sum\limits_{n\neq 0}\frac{%
\varphi _n^\alpha \left( x\right) 
\varphi _n^{\alpha ^{\dagger }}\left(
y\right) }{\lambda _n^\alpha +\epsilon }\textstyle{.} 
$$
In the limit $\epsilon 
$$\rightarrow 0,$ ${\bf S}_\epsilon ^\alpha $ is
regular and can be expressed as \cite{rothe-schroer}, 
\begin{equation}
\label{s-em-fcao-de-g}{\bf S}^\alpha 
\left( x,y\right) =G^\alpha \left(
x,y\right) -\int d^2zG^\alpha 
\left( x,z\right) {\bf P}_0^\alpha \left(
z,y\right) -\int d^2z{\bf P}_0^\alpha 
\left( x,z\right) G^\alpha \left(
z,y\right) , 
\end{equation}
where $G^\alpha \left( x,y\right) $ 
is the fermionic Green function. We can
see that ${\bf S}^\alpha $ satisfies 
\begin{equation}
\label{imposicao a s}D_\alpha {\bf S}^\alpha 
\left( x,y\right) =\delta
\left( x-y\right) -{\bf P}_0^\alpha 
\left( x,y\right) ={\bf S}^\alpha \left(
x,y\right) D_\alpha . 
\end{equation}

In the sector associated with topological 
charge $N$, we shift the fermions
by 
\begin{eqnarray*}
\psi \left( x\right)  &\rightarrow &\psi \left( x\right) -\int d^2y{
\bf{S}}%
\left( x,y\right) \eta \left( y\right) , \\
\overline{\psi }\left( x\right)  &\rightarrow &\overline{\psi }\left(
x\right) -\int d^2y\overline{\eta }\left( y\right) {\bf{S}}\left(
y,x\right) ,
\end{eqnarray*}
where ${\bf S}\left( x,y\right) ={\bf S}^{\alpha =1}\left( 
x,y\right) ,$ to
obtain 
\begin{eqnarray*}
Z\left[ J^\mu ,\overline{\eta },\eta \right]  
&=&\int \left[ dA_\mu \right]
\left[ d\overline{\psi }\right] \left[ d\psi \right] \exp
 \left\langle 
\overline{\eta }\bf{S}\eta \right\rangle  \\
&&\times \exp \left( -S+\left\langle J^\mu A\mu \right\rangle 
+\left\langle 
\overline{\eta }\bf{P}_0\psi \right\rangle +\left\langle \overline{
\psi }%
\bf{P}_0\eta \right\rangle \right) .
\end{eqnarray*}

At this point, it must be stressed 
the role played by the external sources.
They are responsible for preventing 
$Z$ from vanishing, by picking up the
zero modes explicitly.

Now we can bosonize the theory in 
this sector, performing the change of
variables 
\begin{eqnarray*}
\psi  &\rightarrow &\exp \left( -i\rho +\phi \gamma _5\right) \psi ,
\\
\overline{\psi } &\rightarrow &\overline{\psi }\exp \left( i\rho 
+\phi
\gamma _5\right) ,
\end{eqnarray*}
where $\rho $ and $\phi $ were already 
defined in (\ref{ro-fi}). Taking into
account the Fujikawa jacobian, we will end up with \cite{tiao-teresa}
\begin{eqnarray}
Z\left[ J^\mu ,\overline{\eta },\eta \right]  &=&\sum_N\int \left[
 da_\mu
\right] \Im \left( a_\mu ,A_\mu ^{\left( N\right) }\right) \exp 
\left(-
\left\langle \frac 14F_{\mu \nu }F^{\mu \nu }\right\rangle 
+\left\langle
J^\mu A\mu \right\rangle \right)   \label{func-ger} \\
&&\times \exp \left( \left\langle \overline{\eta }^{\prime }\bf{S}%
^{\left( N\right) }\eta ^{\prime }\right\rangle \right) \det {}^{
\prime
}D^{\left( N\right) }\prod_{i=1}^{\left| N\right| }\left\langle 
\overline{\eta }^{\prime }\varphi _{0i}^{\left( N\right) 
}\right\rangle 
\left\langle
\varphi _{0i}^{\left( N\right) ^{\dagger }}\eta ^{\prime 
}\right\rangle , 
\nonumber
\end{eqnarray}
where 
\begin{eqnarray}
\eta ^{\prime } &=&\exp \left( i\rho +\phi \gamma _5\right) \eta 
\textstyle{,}
\label{fontes-linha} \\
\bar{\eta}^{\prime } &=&\bar{\eta}\exp \left( -i\rho 
+\phi \gamma _5\right) 
\textstyle{,}  \nonumber
\end{eqnarray}
and 
$$
\Im \left( a_\mu ,A_\mu ^{\left( N\right) }\right) \equiv \frac{\det
^{\prime }D}{\det ^{\prime }D^{\left( N\right) }}N\left[ \phi \right]
\textstyle{,} 
$$
where $N\left[ \phi \right] $ is given by%
$$
N^{-1}\left[ \phi \right] =\det \left| 
\left\langle \varphi _{0i}^{\left(
N\right) ^{\dagger }}\exp \left( 2\phi \gamma _5\right) \varphi
_{0j}^{\left( N\right) }\right\rangle \right| . 
$$

To compute the ratio of determinants, 
we can make use of the formal relation 
$$
\det D=\exp Tr\ln D, 
$$
where, instead of $\det D,$ we use \cite{gamboa-saravi}

$$
\det {}^{\prime }D_\alpha =\lim _{\epsilon 
\rightarrow 0^{+}}\frac{\det
\left( D_\alpha +\epsilon {\bf 1}\right) }
{\epsilon ^{\left| N\right| }}. 
$$
Now, 
\begin{equation}
\label{det-linha}\frac d{d\alpha }
\det {}^{\prime }D_\alpha =\lim _{\epsilon
\rightarrow 0^{+}}\frac{\det 
\left( D_\alpha +\epsilon {\bf 1}\right) }{%
\epsilon ^{\left| N\right| }}Tr\left[ 
\left( D_\alpha +\epsilon {\ {\bf 1}}%
\right) ^{-1}\frac{dD_\alpha }{d\alpha }\right] \textstyle{,} 
\end{equation}
or, in the limit $\epsilon $$\rightarrow 0^{+}$ 
$$
\frac d{d\alpha }\ln 
\det {}^{\prime }D_\alpha =Tr\left[ {\ {\bf S}}^\alpha 
\frac{dD_\alpha }{d\alpha }\right] . 
$$
Writing $D\equiv D_{\alpha =1}$ and 
$D^{\left( N\right) }\equiv D_{\alpha
=0} $, we do the integration in $\alpha $ to obtain 
\begin{equation}
\label{det-ferm-modos-zero}\ln 
\frac{\det {}^{\prime }D}{\det {}^{\prime
}D^{\left( N\right) }}=\int_0^1d\alpha 
Tr\left[ {\ {\bf S}}^\alpha \frac{%
dD_\alpha }{d\alpha }\right] . 
\end{equation}
The computation of this trace requires the use of some regularization
procedure. Using, for example, the 
point-splitting regularization \cite
{jackiw-johnson}\cite{tiao-cesar-2}, we obtain 
\begin{equation}
\label{ln-dos-dets}\ln 
\frac{\det {}^{\prime }D}{\det {}^{\prime }D^{\left(
N\right) }}=-\Gamma \left[ 
a_\mu \right] -\overline{\Gamma }\left[ A_\mu
^{\left( N\right) },a_\mu \right] +\ln \det \left| 
\left\langle \varphi
_{0i}^{\left( N\right) ^{\dagger }}\exp \left( 2\phi \gamma _5\right)
\varphi _{0j}^{\left( N\right) }\right\rangle \right| , 
\end{equation}
where 
\begin{eqnarray}
\Gamma \left[ a_\mu \right]  &=&\frac{e^2}{4\pi }\int d^2xa_\mu 
\left(
a\left( N\right) \delta _{\mu \nu }-\frac{\partial _\mu 
\partial _\nu }{\Box 
}\right) a_\nu ,  \label{acao efetiva 1} \\
\overline{\Gamma }\left[ A_\mu ^{\left( N\right) },a_\mu \right]  
&=&\frac{e^2}{2\pi }\int d^2xa_\mu \left( a\left( N\right) 
\delta _{\mu \nu }-\frac{\partial _\mu \partial _\nu }{\Box }\right) 
A_\nu ^{\left( N\right) }, 
\nonumber
\end{eqnarray}
and we have written the zero modes of 
$D_\alpha $ in terms of the ones of $%
D^{\left( N\right) }$, 
$$
\varphi _{0i}^\alpha =\exp \left( 
\alpha \left( i\rho +\phi \gamma _5\right)
\right) \sum_{j=1}^ND_{ij}\varphi _{0j}^{\left( N\right) }, 
$$
where the $D_{ij}$ are introduced to 
insure the orthogonality of the $%
\varphi _{0i}^\alpha $. The last term in equation (\ref{ln-dos-dets})
cancels exactly the factor 
$N\left[ \phi \right] $ in the Jacobian. One
should notice the presence of the parameters 
$a\left( N\right) $ in the
results above, that come from the path ordered exponential, put for
occasionally keep gauge invariance ($a=1$). 
We remark that $a\left( N\right) 
$ can be chosen independently 
in each topological sector, what gives an
infinite degree of arbitrariness to the theory.

Actually, it is more convenient to 
express the generating functional in
terms of the original (non orthonormal) 
set of eingenfunctions of $D^{\left(
N\right) },$ obtained by directly solving 
$$
D^{\left( N\right) }\Phi _{0i}^{\left( N\right) }=0, 
$$
where the $\Phi _{0i}^{\left( N\right) }$ , in terms of the 
light-cone
variables, is given by 
\cite{bardakci-crescimano}\cite{manias-naon-trobo} 
\begin{equation}
\label{zero-modes}\Phi _{0i}^{\left( N\right) }=\left\{ 
\begin{array}{c}
z^{i-1}\exp f {1 \choose 0},\qquad i=1,\ldots ,N,N>0 \\ 
\overline{z}^{i-1}\exp \left( -f\right) {0 \choose 1},\qquad 
i=1,\ldots 
,-N,N<0. 
\end{array}
\right. 
\end{equation}

The next step for obtaining the 
generating functional (\ref{func-ger}) is
the computation of 
det$^{\prime }D^{\left( N\right) }$. This can be done if
we use the method presented in \cite{tiao-cesar}, where a functional
differential equation for $\det {}^{\prime }D^{\left( N\right) }$ 
$$
\frac \delta {\delta f\left( 
x\right) }\det {}^{\prime }D^{ \left( N\right)
}=\det {}^{\prime }D^{\left( N\right) }\left[ 
\frac{a\left( N\right) }{2\pi }%
\Box f\left( x\right) +2tr\left( {\bf P}_0^{\left( N\right) }\left(
x,x\right) \gamma _5\right) \right] \textstyle{,} 
$$
can be solved to give 
\begin{equation}
\label{det-linha-N}\det {}^{\prime }D^{\left( 
N\right) }=\exp \left( -\Gamma
^{\prime }\right) \det \left( 
\left\langle \Phi _{0i}^{\left( N\right)
^{\dagger }}\Phi _{0j}^{\left( N\right) }\right\rangle \right) , 
\end{equation}
where 
$$
\Gamma ^{\prime }\left[ A_\mu ^{\left( 
N\right) }\right] =\frac{e^2a\left(
N\right) }{4\pi }\int d^2xf\Box f. 
$$

Finally, if we observe that 
$$
{\bf S}^{\left( N\right) }\left( 
x,y\right) =G^{\left( N\right) }\left(
x,y\right) -\int d^2zG^{\left( 
N\right) }\left( x,z\right) {\ {\bf P}}%
_0^{\left( N\right) }\left( 
z,y\right) -\int d^2z{\bf P}_0^{\left( N\right)
}\left( x,z\right) G^{\left( N\right) }\left( z,y\right) , 
$$
where 
$$
G^{\left( N\right) }\left( x,y\right) =\left\{ 
\exp \left( f\left( x\right)
-f\left( y\right) \right) {\bf P}_{+}+\exp \left( 
-\left( f\left( x\right)
-f\left( y\right) \right) 
\right) {\bf P}_{-}\right\} G_F\left( x,y\right) , 
$$
we conclude, due to the anti-commuting 
nature of $\left\langle \overline{%
\eta }^{\prime }\Phi _{0i}^{\left( N\right) }\right\rangle $ and $%
\left\langle \Phi _{0i}^{\left( N\right) ^{\dagger }}\eta ^{\prime
}\right\rangle $, that we can write 
$$
\exp \left\langle 
\overline{\eta }^{\prime }{\bf S}^{\left( N\right) }\eta
^{\prime }\right\rangle \prod_{i=1}^{\left| 
N\right| }\left\langle \overline{%
\eta }^{\prime }\Phi _{0i}^{\left( 
N\right) }\right\rangle \left\langle \Phi
_{0i}^{\left( N\right) ^{\dagger }}\eta ^{\prime }\right\rangle =\exp
\left\langle \overline{\eta }^{\prime }G^{\left( 
N\right) }\eta ^{\prime
}\right\rangle \prod_{i=1}^{\left| 
N\right| }\left\langle \overline{ \eta }%
^{\prime }\Phi _{0i}^{\left( 
N\right) }\right\rangle \left\langle \Phi
_{0i}^{\left( 
N\right) ^{\dagger }}\eta ^{\prime }\right\rangle \textstyle{,}
$$
giving for $Z$ the expression 
\begin{equation}
\label{func-ger-final}Z\left[ J^\mu ,\overline{\eta },\eta \right]
=\sum_N\int \left[ da_\mu \right] \exp \left( 
-\overline{S} +\left\langle
J^\mu A\mu \right\rangle +\left\langle 
\overline{\eta }^{ \prime }G^{\left(
N\right) }\eta ^{\prime }\right\rangle 
\right) \prod_{i=1}^{ \left| N\right|
}\left\langle \overline{\eta }^{\prime }\Phi _{0i}^{\left( N\right)
}\right\rangle \left\langle \Phi _{0i}^{\left( 
N\right) ^{\dagger }}\eta
^{\prime }\right\rangle , 
\end{equation}
where 
$$
\overline{S}=\left\langle 
\frac 14F_{\mu \nu }F^{\mu \nu }\right\rangle
+\Gamma \left[ a_\mu \right] +\overline{\Gamma }\left[ 
a_\mu ,A_\mu ^{\left(
N\right) }\right] +\Gamma ^{\prime }\left[ 
A_\mu ^{\left( N\right) }\right]
. 
$$

It is important to stress here that this formula for the generating
functional is independent of the 
choice of the representative $f\left(
x\right) ,$ that is 
$$
\frac{\delta Z}{\delta f\left( x\right) }=0. 
$$

\section{Non Trivial Contributions to Correlation Functions}

Being directly proportional to fermionic 
sources, it is not difficult to see
that there are no contributions to bosonic 
correlation functions and that
bosonic-fermionic ones do not give 
different information (concerning the
ambiguities) than that given by the 
fermionic functions alone. We have non
vanishing contributions from non trivial 
topologies to fermionic correlation
functions of the type 
$$
\left\langle \prod_{i=1}^k\psi _{\alpha _i} \left( 
x_i\right) \prod_{ j=1}^k{%
\ \overline{\psi}}_{\beta _j} \left( y_j\right) \right\rangle =\frac
1{Z\left[ 0\right] }\frac 
\delta {\delta \overline{\eta }_{\alpha _1}\left(
x_1\right) }\cdots \frac 
\delta {\delta \eta _{\beta _k}\left( y_k\right)
}\left. Z\left[ 
0,\overline{\eta },\eta \right] \right| _{\overline{\eta }%
=\eta =0}. 
$$
It can be easily shown, by induction, that 
\begin{equation}
\label{integrando-derivado
}\left\langle \prod_{i=1}^k\psi _{\alpha
_i}\left( x_i\right) \prod_{j=1}^k{\overline{\psi}}_{\beta _j}\left(
y_j\right) \right\rangle =\sum_{k=-N }^N\int \left[ 
da_\mu \right] \exp
\left( -\bar S\right) \det \left| 
\begin{array}{cc}
{\bf \Phi }^{\prime \left( 
N\right) ^{\dagger }} & {\bf \emptyset } \\ {\bf G%
}^{\prime \left( N\right) } & {\bf \Phi }^{\prime \left( N\right) } 
\end{array}
\right| , 
\end{equation}
where ${\bf \Phi }$$^{\prime \left( 
N\right) ^{\dagger }}$ is a $N\times k$
matrix given by 
$$
{\bf \Phi }^{\prime \left( N\right) ^{\dagger }}=\left( 
\begin{array}{c}
\Phi _{01}^{\prime \left( N\right) ^{\dagger }}\left( 
y_1\right) \cdots \Phi
_{01}^{\prime \left( N\right) ^{\dagger }}\left( y_k\right) \\ 
\vdots \\ 
\Phi _{0N}^{\prime \left( N\right) ^{\dagger }}\left( 
y_1\right) \cdots \Phi
_{0N}^{\prime \left( N\right) ^{\dagger }}\left( y_k\right) 
\end{array}
\right) , 
$$
${\bf \Phi }$$^{\prime \left( N\right) }$ is $k\times N$, 
$$
{\bf \Phi }^{\prime \left( N\right) }=\left( 
\begin{array}{c}
\Phi _{01}^{\prime \left( N\right) }\left( x_1\right) \cdots \Phi
_{0N}^{\prime \left( N\right) }\left( x_1\right) \\ 
\vdots \\ 
\Phi _{01}^{\prime \left( N\right) }\left( x_k\right) \cdots \Phi
_{0N}^{\prime \left( N\right) }\left( x_k\right) 
\end{array}
\right) , 
$$
and 
$$
{\bf G}^{\prime \left( N\right) }=\left( 
\begin{array}{c}
G^{\prime \left( N\right) }\left( 
x_1,y_1\right) \cdots G^{\prime \left(
N\right) }\left( x_1,y_k\right) \\ 
\vdots \\ 
G^{\prime \left( N\right) }\left( 
x_k,y_1\right) \cdots G^{\prime \left(
N\right) }\left( x_k,y_k\right) 
\end{array}
\right) 
$$
and ${\bf \emptyset }$ (the null matrix) 
are square matrices $k\times k$ and 
$N\times N$ respectively, and 
\begin{eqnarray*}
G^{^{\prime
 }\left( N\right) }\left( x_i,y_j\right)  &=&\exp \left( -i\rho
+\phi \gamma _5\right) G^{\left( N\right)
 }\left( x_i,y_j\right) \left(
i\rho +\phi \gamma _5\right) , \\
\Phi _{0i}^{^{\prime
 }\left( N\right) }\left( x_j\right)  &=&\exp \left(
-i\rho +\phi \gamma _5\right) \Phi _{0i
}^{\left( N\right) }\left( x_j\right)
, \\
\Phi _{0i}^{^{\prime 
}\left( N\right) ^{\dagger }}\left( y_j\right)  &=&\Phi
_{0i}^{\left( N\right) ^{\dagger 
}}\left( y_j\right) \exp \left( i\rho +\phi
\gamma _5\right) .
\end{eqnarray*}
If we define 
$$
\chi _{ij}=\left\{ 
\begin{array}{c}
\left( z_j\right) ^{i-1}{1 \choose 0},\qquad N>0 \\ 
\left( \bar z_j\right) ^{i-1}{0 \choose 1},\qquad N<0 
\end{array}
\right. 
$$
and use the expressions for the zero 
modes (\ref{zero-modes}) we can show
that 
\begin{eqnarray*}
\det \left( \bf{\Phi }^{^{\prime }\left( N\right) }\right)  &=&\exp
\left( \sum_{i=1}^{\left| N\right|
 }i\rho \left( y_i\right) \right) \exp
\left( \pm \sum_{i=1}^{\left| N\right| 
}f\left( y_i\right) +\phi \left(
y_i\right) \right) \det \left( \bf{\chi }\right) , \\
\det \left( \bf{\Phi }^{^{\prime }\left( N\right) ^{\dagger 
}}\right) 
&=&\exp \left( -\sum_{i=1}^{\left| N\right| 
}i\rho \left( x_i\right) \right)
\exp \left( \pm \sum_{i=1}^{\left| N\right|
 }f\left( x_i\right) +\phi \left(
x_i\right) \right) \det \left( \bf{\chi }^{\dagger }\right) 
\end{eqnarray*}
and 
\begin{eqnarray*}
\det \left( \bf{G}^{\prime \left( N\right) }\right)  &=&\exp \left(
\sum_{i=\left| N\right| +1
}^ki\left( \rho \left( y_i\right) -\rho \left(
x_i\right) \right) \right) \times  \\
&&\exp \left( \pm \sum_{i=\left| N\right| +1
}^kf\left( y_i\right) -f\left(
x_i\right) 
+\phi \left( y_i\right) -\phi \left( x_i\right) \right) \det
\left( \bf{G}_F\right) ,
\end{eqnarray*}
where, according to the positiveness or not of $N$, 
\begin{eqnarray*}
\det \left( \bf{\chi 
}\right)  &=&\prod^{|N|}_{{\stackrel {i,j=1}{i>j} }}
\left| z_i-z_j\right| \otimes {1 \choose 0},\qquad N>0 \\
\det \left( \bf{\chi }^{\dagger }\right)  &=&\prod^{|N|}_{
{\stackrel {i,j=1}{i>j} }}
\left| \bar{z}_i-\bar{z}_j\right| \otimes \left( 
\begin{array}{cc}
1 & 0
\end{array}
\right) ,\qquad N>0
\end{eqnarray*}
or 
\begin{eqnarray*}
\det \left( \bf{\chi 
}\right)  &=&\prod^{|N|}_{{\stackrel {i,j=1}{i>j} }}
\left| \bar{z}_i-\bar{z
}_j\right| \otimes {0 \choose 1},\qquad N<0 \\
\det \left( \bf{\chi }^{\dagger }\right)  &=&\prod^{|N|}_{
{\stackrel {i,j=1}{i>j} }}
\left| z_i-z_j\right| \otimes \left( 
\begin{array}{cc}
0 & 1
\end{array}
\right) ,\qquad N<0.
\end{eqnarray*}

Collecting these results, we arrive at 
\begin{equation}
\label{func-corr-final
}\left\langle \prod_{i=1}^k\psi _{\alpha _i}\left(
x_i\right) \prod_{j=1}^k{\overline{\psi }}_{\beta _j}\left( 
y_j\right)
\right\rangle =\sum_{k=-N}^N\int \left[ 
da_\mu \right] \exp \left( -\bar S_{%
\mbox {\tiny {\rm sources}}}\right) \det \left| 
\begin{array}{cc}
{\bf \chi }^{\dagger } & {\bf \emptyset } \\ {\bf G}_F & {\bf \chi } 
\end{array}
\right| \textstyle{,} 
\end{equation}
where 
$$
\bar S_{\mbox  {\tiny {\rm sources}}}=\bar S-\left\langle i\left( 
j_\rho
+j_\rho ^{\prime }\right) \rho \right\rangle \mp \left\langle \left(
j+j^{\prime }\right) \left( f+\phi \right) \right\rangle , 
$$
the signs $\mp $ refer to $N>0$ and $N<0$ respectively, 
and $j_\rho $, $%
j_\rho ^{\prime }$, $j$ and $j^{\prime }$ are defined by 
\begin{eqnarray*}
j_\rho  &=&\sum_{i=1}^{\left| N\right| 
}\delta \left( y_i-z\right) -\delta
\left( x_i-z\right) , \\
j_\rho ^{\prime } &=&\sum_{i=\left| N\right| +1
}^k\delta \left( y_i-z\right)
-\delta \left( x_i-z\right) , \\
j  &=&\sum_{i=1}^{\left| N\right| 
}\delta \left( y_i-z\right) +\delta
\left( x_i-z\right) , \\
j^{\prime } &=&\sum_{i=\left| N\right| +1
}^k\delta \left( y_i-z\right)
-\delta \left( x_i-z\right) ,
\end{eqnarray*}
with $\left\langle {\ }\right\rangle $ 
representing integration over $z.$

There is still a last integration over the 
scalar fields $\rho $ and $\phi $
in terms of which the gauge field is written. 
So we write the effective
action $\overline{S}$ in terms of these fields

$$
\overline{S}=\frac 1{2e^2}\left\langle \left( 
\phi +f\right) \Box \left(
\Box -\frac{e^2a\left( N\right) }\pi \right) \left( \phi +f\right)
\right\rangle -\frac{\left( a\left( 
N\right) -1\right) }{2\pi }\left\langle
\rho \Box \rho \right\rangle \textstyle{,} 
$$
and do the following change of variables 
using the sources $j_\rho ,$ $%
j_\rho ^{\prime },$ $j$ and $j^{\prime }$: 
\begin{equation}
\label{mudanca-ro
}\sigma =\rho +\frac i\lambda \left\langle \Delta
_F\left( j_\rho +j_\rho ^{\prime }\right) \right\rangle 
\end{equation}
and 
\begin{equation}
\label{mudanca-f-fi}\varphi =\phi +f\mp e^2\left\langle \Delta \left(
m;x-y\right) \left( j+j^{\prime }\right) \right\rangle 
\end{equation}
where 
$$
\Delta _F\left( x-y\right) \equiv \Box ^{-1}\left( 
x-y\right) =\frac 1{2\pi
}\ln m\left| x-y\right|  
$$
and 
$$
\Delta \left( m;x-y\right) \equiv \left[ \Box \left( \Box -m^2\right)
\right] ^{-1}\left( x-y\right) =-\frac 1{2\pi m^2}\left\{ 
K_0\left[ m\left|
x-y\right| \right] +\ln m\left| x-y\right| \right\}  
$$
($K_0$ is the zeroth-order modified Bessel function) 
and we have defined $%
\lambda \equiv \left( a\left( 
N\right) -1\right) /\pi $ and $m^2=\left(
e^2a\left( 
N\right) \right) /\pi .$ Now we have for $\overline{S}$ plus the
sources the expression 
\begin{eqnarray*}
\overline{S}_{\mbox {\tiny {\rm sources}}} &=&\frac 1{2e^2
}\left\langle \varphi \Box \left(
\Box -m^2 \right) \varphi \right\rangle 
-\frac{e^2}2\left\langle \left( j+j^{\prime
 }\right) \Delta \left( m\right) 
\left(
j+j^{\prime }\right) \right\rangle - \\
&&\frac{\lambda }{2 
}\left\langle \sigma \Box
\sigma \right\rangle -\frac 1{2\lambda 
}\left\langle \left( j_\rho +j_\rho
^{\prime }\right) \Delta _F\left( j_\rho +j_\rho ^{\prime }\right)
\right\rangle .
\end{eqnarray*}

As we have already said, the scalar fields 
$\rho $ and $\phi $ are such that 
$a_\mu $ does not carry a topological 
charge in the limit $\left| x\right|
\rightarrow \infty .$ So it is desirable 
that the new fields $\sigma $ and $%
\varphi $ behave like the old ones, going to zero at infinity. 
If this would
not be the case, it would be equivalent 
to perform transformations that
change the topological sector, 
which would lead us to compute jacobians over
noncompact spaces, what is very difficult to obtain \cite
{manias-naon-trobo-2}\cite{gamboa}. 
So, altough keeping in mind the general
case, we will restrict ourselves to 
transformations which do not change the
topological sector.

In the case of the $\sigma $ field we have 
\begin{eqnarray*}
\lim_{\left| x\right| \rightarrow \infty }\sigma \left( x\right) 
&=&\lim_{\left| x\right| \rightarrow \infty 
}\rho \left( x\right) +\frac
i\lambda \lim_{\left| x\right| \rightarrow \infty 
}\left\langle \Delta
_F\left( x-z\right) \left( j_\rho +j_\rho ^{\prime
 }\right) \right\rangle  \\
&=&\frac {i}{2\pi \lambda }\lim_{\left| x\right| \rightarrow \infty
 }\left\{
\sum_{i=1
}^k\left( \ln m\left| x-x_i\right| -\ln m\left| x-y_i\right| \right)
\right\}  \\
&=&0,
\end{eqnarray*}
once $\lim _{\left| x\right| \rightarrow \infty }\rho \left( 
x\right) =0,$
in agreement with the conditions imposed.

For the field $\varphi ,$ we have 
\begin{eqnarray*}
\lim_{\left| x\right| \rightarrow \infty }\varphi \left( x\right) 
&=&\lim_{\left| x\right| \rightarrow \infty 
}f\left( x\right) +\lim_{\left|
x\right| \rightarrow \infty }\phi \left( x\right) \mp \lim_{
\left| x\right|
\rightarrow \infty 
}e^2\left\langle \Delta \left( m;x-z\right) \left(
j+j^{\prime }\right) \right\rangle  \\
&=&-N\ln \left| x\right| \pm \frac{e^2}{2\pi m^2}\lim_{
\left| x\right|
\rightarrow \infty }\left\langle \left( K_0\left[ m\left|
x-z\right| \right] +\ln m\left| x-z\right| \right) \left( j+j^{\prime
}\right) \right\rangle  \\
&=&-N\ln \left| x\right| \pm \frac 1{2a\left( N\right) 
}2\left| N\right| \ln
\left| x\right|  \\
&=&-\left( N\mp \frac{\left| N\right| }{a\left( N\right) 
}\right) \ln \left|
x\right| ,
\end{eqnarray*}
once $K_0$ is well behaved in the limit considered and $\lim _{\left|
x\right| \rightarrow \infty }\phi \left( 
x\right) =0$. Here, the $\mp $ sign
corresponds to sectors with topological 
charge $N$ and $-N,$ respectivelly.
The assymptotic behavior of $\varphi $ is then 
$$
\lim _{\left| x\right| \rightarrow \infty }\varphi \left( 
x\right) =\left\{ 
\begin{array}{c}
-\left( N-\frac N{a\left( 
N\right) }\right) \ln \left| x\right| ,\qquad N>0,
\\ 
-\left( N-\frac N{a\left( 
N\right) }\right) \ln \left| x\right| ,\qquad N<0. 
\end{array}
\right. 
$$
which is singular unless we have 
$$
a\left( N\right) =1,\qquad \forall \quad N\neq 0. 
$$

\section{Correlation functions in sectors with trivial topology}

In this case, bosonization gives us 
\begin{equation}
\label{func-ger-trivial}Z\left[ 
J^\mu ,\overline{\eta },\eta \right] =\int
\left[ dA_\mu \right] \Im \left( 
A_\mu \right) \exp \left( -\left\langle
\frac 14F_{\mu \nu }F^{\mu \nu }\right\rangle +\left\langle 
J^\mu A_\mu
\right\rangle +\left\langle \overline{\eta }^{\prime }\left( 
i\gamma ^\mu
\partial _\mu \right) ^{-1}\eta ^{\prime }\right\rangle \right) , 
\end{equation}
where $\overline{\eta }^{\prime }$ and $\eta ^{\prime }$ 
are the transformed
fermionic sources (\ref{fontes-linha}), and

$$
\Im \left( 
A_\mu \right) =\frac{\det D}{\det i\gamma ^\mu \partial _\mu }%
\textstyle{,} 
$$
can be written as an effective term to be added to the action 
$$
\overline{S}=\left\langle 
\frac 14F_{\mu \nu }F^{\mu \nu }\right\rangle
+\left\langle \frac{e^2}{2\pi }A_\mu \left\{ 
a\delta _{\mu \nu }-\frac{%
\partial _\mu \partial _\nu }{\Box }\right\} A_\nu \right\rangle . 
$$

To compute the photon self-energy 
we consider the fermionic external sources
to be absent $\overline{\eta }=\eta =0$, and write 
\begin{eqnarray*}
\left\langle J_\mu A^\mu \right\rangle &=&\left\langle 
J_\mu \left( \partial
^\mu \rho +\tilde{\partial}^\mu \phi \right) \right\rangle \\
&=&-\left\langle \left( \partial _\mu J^\mu \right) \rho 
+\left( \tilde{\partial}_\mu J^\mu \right) \phi \right\rangle \\
&=&-\left\langle J_L\rho +J_T\phi \right\rangle ,
\end{eqnarray*}
where $J_L$ and $J_T$ are the 
sources associated to the longitudinal and
transverse terms, respectivelly. Now the action goes into 
\begin{equation}
\label{seff-fontes}\overline{S }-
\left\langle J_\mu A^\mu \right\rangle
=\int d^2x\left\{ \frac 1{2e^2 }\phi \Box \left( 
\Box -m^2\right) \phi
-\frac \lambda 2\rho \Box \rho +J_L\rho +J_T\phi \right\} 
\textstyle{.} 
\end{equation}

We perform a change of variables 
on the scalar fields $\phi $ and $\rho $ to
obtain the following gaussian expression 
for $Z\left[ J^\mu \right] :$%
$$
Z\left[ J_\mu \right] =\exp \left\langle \frac{e^2 }2J_T\left[ 
\Box \left(
\Box -m^2\right) \right] ^{-1 }J_T\right\rangle 
\exp \left\langle -\frac
1{2\lambda }J_L\Box ^{-1}J_L\right\rangle \textstyle{.} 
$$

In momentum space, this is simply

$$
Z\left[ J_\mu \right] =\exp \left\{ \int \frac{d^2k}{\left( 
2\pi \right) ^2}%
\tilde J_\mu \left( -k\right) \left[ 
\frac{e^2}2\frac{k^2\delta ^{ \mu \nu
}-k^\mu k^\nu }{k^2\left( 
k^2+m^2\right) }-\frac 1{2\lambda }\frac{ k_\mu
k_\nu }{k^2}\right] \tilde J_\nu \left( k\right) \right\} , 
$$
which gives to the photon self-energy the expression 
$$
G^{\mu \nu }\left( x-y\right) =\int \frac{d^2p}{ 
\left( 2\pi \right) ^2}\exp
\left( -ip\left( x-y\right) \right) \left[ 
\frac{e^2}2\frac{ p^2\delta ^{\mu
\nu }-p^\mu p^\nu }{p^2\left( 
p^2+m^2\right) }-\frac 1{2\lambda }\frac{p_\mu
p_\nu }{p^2}\right] . 
$$

We see that the use of the point-splitting 
regularization in the computation
of the fermionic determinant introduced 
the Jackiw parameter explicitly at
the physical pole of the propagator. 
This implies that the ambiguity will be present
in any physical quantity we compute, which depends on this pole.

One can easily see, through the expression above or the next, that
$$
G^{\mu \nu }\left( x-y\right) =-e^2\left( 
\Box \delta ^{\mu \nu }-\partial
^\mu \partial ^\nu \right) \left[ 
2\pi \Delta \left( m;x-y\right) \right]
+\frac 1{2\pi \lambda }\partial ^\mu 
\partial ^\nu \left[ \ln \left( m\left|
x-y\right| \right) \right] 
$$
that $G^{\mu \nu }$ is free of ultraviolet and infrared divergences.

In the case of the fermionic self-energy, we can consider the bosonic
external source to be absent instead, $J^\mu =0$. We define 
\begin{eqnarray*}
S_{\alpha \beta }^{\pm }\left( x-y\right) _{J^\mu =0
} &\equiv &\left( \left(
i\gamma ^\mu \partial _\mu \right) ^{-1}\bf{P}_{\pm }\right) _{
\alpha
\beta } \\
&&\int dA_\mu \exp \left( -\overline{S
}\right) \exp \left( i\left( \rho
\left( x\right) -\rho \left( y\right) \right) \mp \left( \phi \left(
x\right) -\phi \left( y\right) \right) \right) \textstyle{,}
\end{eqnarray*}
which allow us to write 
$$
S_{\alpha \beta }\left( x-y\right) =S_{\alpha \beta }^{+}\left( 
x-y\right)
+S_{\alpha \beta }^{-}\left( x-y\right) \textstyle{.} 
$$
Similar calculations as before give 
$$
S_{\alpha \beta }^{\pm }\left( x-y\right) _{J^\mu =0}=\left( \left( 
i\gamma
^\mu \partial _\mu \right) ^{-1}
{\bf P}_{\pm }\right) _{\alpha \beta }\exp
\left\langle \frac{e^2}2j\left[ \Box \left( \Box -m^2\right) \right]
^{-1}j+\frac 1{2\lambda }j\Box ^{-1}j\right\rangle \textstyle{,} 
$$
where $j\left( z\right) =\delta \left( 
x-z\right) -\delta \left( y-z\right) $%
. Remembering that ${\bf P}_{+}+{\bf P}_{-}={\bf 1},$ we find 
\begin{equation}
\label{tiao1}S_{\alpha \beta }\left( 
x-y\right) =\exp \left\langle \frac{e^2}%
2j\left[ \Box \left( 
\Box -m^2\right) \right] ^{-1}j+\frac 1{2\lambda }j\Box
^{-1}j\right\rangle S_{\alpha \beta }^F\left( x-y\right) , 
\end{equation}
where $S_{\alpha \beta }^F\left( 
x-y\right) $ is the two point function of
free fermions.

About the ultraviolet behavior 
we see that, being $S_{\alpha \beta }^F$
finite, the expression 
\begin{eqnarray}
\left\langle \frac{e^2}2j\left[ \Box \left( \Box -m^2\right) \right]
^{-1}j\right\rangle &=&e^2\int \frac{d^2k}{\left( 2\pi 
\right) ^2}\frac{1-\exp \left( ik\left( x-y\right) \right) }{
k^2\left( 
k^2+m^2\right) }
\label{regula 1} \\
&=&e^2\left( 2\pi \Delta\left( m;0\right) \right) -e^2\left( 2\pi 
\Delta\left( m;x-y \right) \right) ,  \nonumber
\end{eqnarray}
is free of singularities, while 
\begin{eqnarray}
\left\langle \frac 1{2\lambda }j\Box ^{-1
}j\right\rangle &=&\lim_{\left| 
\alpha \right| \rightarrow 0}\frac 1{2\pi \lambda 
}\left[ \ln \left(
m\left| \alpha \right| \right) -\ln 
\left( m\left| x-y\right| \right) \right]  \label{regula 2} \\
&=&\lim_{\alpha \rightarrow 0}\frac 1{2\pi \lambda 
}\ln \left( \frac {\left| \alpha \right| }
{\left| x-y\right| }\right)  \nonumber \\
&=&\left\{ 
\begin{array}{c}
-\infty ,\qquad {\mbox {\rm { if }}}\lambda >0 \\ 
+\infty ,\qquad {\mbox {\rm { if }}}\lambda <0
\end{array}
\right. 
\nonumber
\end{eqnarray}
give us for (\ref{tiao1}) 
$$
S_{\alpha \beta }\left( x-y\right) =\left\{ 
\begin{array}{c}
0,\qquad 
{\mbox {\rm { if }}}\lambda >0 \\ \infty ,\qquad 
\negthinspace \negthinspace 
\negthinspace {\mbox {\rm { if }}}\lambda <0 
\end{array}
\right. , 
$$
which, in the case $\lambda <0$, 
has a divergence that depends explicitly on
the ambiguity.

In obtaining this result we have used 
both infrared ($m$) and ultraviolet ($%
\alpha $) regulators. The infrared 
divergence cancels but the ultraviolet
one remains. It is important to stress 
the fact that for $\lambda =0$, we do
not have any divergences.

For $\lambda <0$, we can perform a wave function renormalization 
\begin{eqnarray}
\psi &=& Z_\psi ^{\frac 12}\psi ^{\prime },  \label{renormalizacao}
\\
\overline{\psi } &=& Z_{\overline{\psi }}^{\frac 12}\overline{\psi }%
^{\prime },  \nonumber
\end{eqnarray}
which gives 
$$
S_{\alpha \beta }^R\left( 
x-y\right) =Z_\psi ^{\frac 12}Z_{\overline{\psi }%
}^{\frac 12}S_{\alpha \beta }\left( x-y\right) . 
$$
By choosing 
$Z_\psi ^{\frac 12}=Z_{\overline{\psi }}^{\frac 12}=Z^{\frac
12}, $ we will need just one 
renormalization condition to fix $Z^{\frac 12},$
for example, 
$$
Z^{\frac 12}=\exp \left\{ 
-\frac 1{2\pi \lambda }\ln \left| \alpha \right|
+\beta \right\} , 
$$
where $\beta $ is an additional 
ambiguity over the finite part of $Z^{\frac
12}$ to be fixed by the requirement 
of external renormalization conditions.
This will lead to a finite result at the end.

To proceed, we compute the mixed four point function 
$$
G_{\alpha \beta }^{\mu \nu }\left( 
x,y,z,w\right) =\frac{\delta ^2 }{\delta
J_\mu \left( x\right) \delta J_\nu \left( 
y\right) }\left\{ T_{ \alpha \beta
}^{+}\left[ J_\mu ;z-w\right] +T_{\alpha \beta }^{- }\left[ J_\mu
;z-w\right] \right\} , 
$$
where

$$
T_{\alpha \beta }^{\pm }\left[ J_\mu ;z-w\right] =\left( 
\left( i\gamma ^\mu
\partial _\mu \right) ^{-1}\left( 
z-w\right) {\bf P}_{\pm }\right) _{\alpha
\beta }\int \left[ dA_\mu \right] \exp \left( -S^{\pm }\right) , 
$$
and 
\begin{eqnarray*}
S^{\pm } &=&\overline{S}-\frac{e^2
}2\left\langle j\left( \left[ \Box \left(
\Box -m^2\right) \right] ^{-1}+\frac 1{e^2\lambda }\Box ^{-1}\right)
j\right\rangle \mp \frac{e^2
}2\left\langle J_T\left[ \Box \left( \Box
-m^2\right) \right] ^{-1}j\right\rangle  \\
&&\mp \frac{e^2
}2\left\langle j\left[ \Box \left( \Box -m^2\right) \right]
^{-1}J_T\right\rangle -\frac i{2\lambda }\left\langle J_L\Box
^{-1}j\right\rangle -\frac i{2\lambda }\left\langle j\Box
^{-1}J_L\right\rangle .
\end{eqnarray*}
We still have to compute functional 
derivatives with respect to $J_\mu $ of
the above expression, which is in the 
form $\exp \left( -\left\langle
JKJ\right\rangle -\left\langle JL\right\rangle \right) .$ 
When taking the
limit $J=0,$ we find 
$$
\frac \delta {\delta J_x\delta J_y}\exp \left( -\left\langle
JKJ\right\rangle -\left\langle JL\right\rangle \right)
_{J_x=J_y=0}=-2K_{xy}+L_xL_y, 
$$
where $K_{xy }$ stands for the bosonic two point function, already
calculated and 
$$
L_{\pm }^\mu \left( z\right) =\int d^2z^{\prime } \left( 
\pm \frac{e^2 }%
2\epsilon ^{\mu \nu }\partial _\nu ^z\left[ \Box \left( 
\Box -m^2\right)
\right] ^{ -1}\left( z,z^{\prime }\right) j\left( 
z^{\prime }\right) +\frac
i{2\lambda }\partial _z^\mu \Box ^{-1}\left( 
z,z^{\prime }\right) j\left(
z^{\prime }\right) \right) . 
$$
Putting all together, we can write 
\begin{eqnarray*}
G_{\alpha \beta }^{\mu \nu }\left( x,y,z,w\right)  &=&S_{\alpha \beta
}\left( z-w\right) G^{\mu \nu }\left( x-y\right) +S_{\alpha \beta
}^{+}\left( z-w\right) L_{+}^\mu \left( x\right) L_{+
}^\nu \left( y\right) 
\\
&&+S_{\alpha \beta }^{-}\left( z-w\right) L_{-}^\mu \left( x\right)
L_{-}^\nu \left( y\right) 
\end{eqnarray*}
where $L_{\pm }^\mu \left( x\right) \equiv L_{\pm }^\mu \left( 
x,z,w\right) $
and $L_{\pm }^\mu \left( y\right) \equiv L_{\pm }^\mu \left( 
y,z,w\right) $.

We see that $S_{\alpha \beta }^{\pm }$ 
has the same ultraviolet behaviour as 
$S_{\alpha \beta }.$ $L_{\pm }^\mu $ is given by 
\begin{eqnarray*}
L_{\pm }^\mu \left( x,z,w\right) &=&\mp \pi e^2\varepsilon _{\mu \nu
}\partial ^\nu \left[ \Delta \left( m;x-z \right) 
+\Delta \left( m;x-w \right) \right] + \\
&&\frac i{4\pi \lambda 
}\partial _\mu \left[ \ln \left( m\left| x-z\right|
\right) +\ln \left( m\left| x-w\right| \right) \right] ,
\end{eqnarray*}
which is free of singularities. 
This shows that the four point function will
be finite and non vanishing if 
the fermionic two point function is. This
analysis can be extended to correlation 
functions with arbitrary number of
legs and the same conclusion will be 
reached, i.e., the only divergence to
be regulated is that of the two point fermionic function.

\section{Conclusion and remarks}

The ambiguity in the Jackiw parameter 
can now be restricted, in the case of
trivial topology. We have found a renormalization for the fermionic
self-energy. This means that the 
theory can be made finite, unitary and
having non vanishing fermionic correlation functions for 
$$
\lambda <0\Longrightarrow \frac{a-1}\pi <0, 
$$
or 
$$
a<1. 
$$
For $a>1$, every correlation function 
involving fermions will vanish, thus
giving an inconsistent theory in the sense that we begin considering
fermionic operators and find, at the end, that these operators are
identically null. For these values of $a$, 
the longitudinal components of $%
A_\mu $ behave like ghosts, thus spoiling 
unitarity, aas can be easily seen
from equation (\ref{seff-fontes}), for example. 

On the other side, we can restrict even 
more the values of $a$ if we do not
admit a tachyon in the spectrum. This 
extra consideration puts the ambiguity
in the interval 
$$
0\leq a\leq 1\textstyle{.} 
$$

We can interpret this range if we 
compare our results with those obtained
after the computation of the conservation 
of the gauge and axial currents in
the Schwinger model given by 
$$
\left\langle \partial _\mu J_5^\mu 
\right\rangle =-\frac e{4\pi }\epsilon
^{\mu \nu }F_{\mu \nu }\left( 
1+a_s\right) =-\frac e{2\pi }\epsilon ^{\mu
\nu }F_{\mu \nu }a 
$$
and 
$$
\left\langle \partial _\mu J^\mu 
\right\rangle =\frac e{2\pi }\partial _\mu
A^\mu \left( 1-a_s\right) =\frac e\pi 
\partial _\mu A^\mu \left( 1-a\right) 
$$
where, $a_s=\left( 1+a\right) /2$ is 
the original parameter introduced by
Jackiw and Johnson \cite{jackiw-johnson}. 
We see that the $a$ parameter
interpolates continuously beetwen 
regularization schemes that preserve
chiral ($a=0$) and gauge ($a=1$) invariances respectively.

At the same time, the divergence found 
for all values of $a\neq 1$ means
that perhaps the ambiguity is only apparent. 
As is well known, whenever we
have to renormalize a theory, we are 
forced to fix our renormalization
counterterms through the use of 
renormalization conditions. These conditions
usually introduce an arbitrary parameter $\mu $ in the correlation
functions, but for a physical quantity $R$ 
one can always prove that \cite
{gross} 
$$
\frac d{d\mu }R=0\textstyle{.} 
$$
This is equally valid for the physical 
masses of the model and simply means
that once we have fixed the experimental 
values of the parameters which
enter into the lagrangean, it does not 
matter the way one chooses to
renormalize the theory. In this way, we 
intend to investigate a possible
dependence of $a$ on renormalization 
group parameters, through a careful
study of the renormalization conditions, 
in a nonperturbative setting. The
main problem that we have to face is 
how to express these conditions
directly in configuration space, 
instead of momentum space, where
bosonization is rather involved. Progress in 
this direction will be reported
elsewhere.

Finally, we would like to remark that, 
in nontrivial topology sectors, the
question seems to be even more difficult 
to answer. As we have seen, there
is an infinite amount of ambiguity in 
the theory, due to arbitrary choices
of $a\left( N\right) $ for each $N$. 
A simple criterium to choose $a\left(
N\right) $ seems to be the one which 
does not allow changes in the
topological sector. It gives a value for 
$a\left( N\right) $ which coincides
with the one obtained through the 
requirement of gauge invariance. The
connection between gauge invariance and 
preservation of topology is not
completely clear, as it sugests the 
existence of some relation between short
and long distance properties of the 
theory, and perhaps can only be
clarified if one could compute the 
correlation functions without these
criteria. It is our aim to explore also 
this direction in the near future.

{\bf Acknowledgements:} S. A. Dias wishes 
to dedicate this work, with love,
to his wife Ana C\'\i ntia. M. B. Silva Neto 
wishes to thank CNPq and CAPES
for financial support.

\end{document}